\documentclass[letter,center,fleqn]{amsart}
\usepackage{amsaddr}
\usepackage{xspace}
\usepackage{url}
\usepackage{graphicx}
\usepackage{amsmath}
\usepackage{amssymb}
\usepackage{caption}
\usepackage{subcaption}
\usepackage[hidelinks]{hyperref} 
\usepackage{color} 

\newcommand{\seq}{{\bf s}\xspace}
\newcommand{\aseq}{{\bf a}\xspace}

\def\footnotesize{\Tiny}

\title[{\Tiny RNAiFold 2.0: A W.S. and software to design custom and Rfam-based RNA molecules}]{RNAiFold 2.0: A web server and software to design custom and 
Rfam-based RNA molecules}
\author{Juan Antonio Garcia-Martin\,$^{1}$,
Ivan Dotu\,$^{2,}$,
Peter Clote\,$^{1\dag}$
}
\address{%
\small
$^{1}$Biology Department, Boston College, 140 Commonwealth Avenue,
Chestnut Hill, MA 02467 (USA) \\
$^{2}$Research Programme on Biomedical Informatics (GRIB), 
Department of Experimental and Health Sciences, Universitat Pompeu Fabra, 
IMIM (Hospital del Mar Medical Research Institute);
Dr. Aiguader, 88. Barcelona (Spain) \\
$^{\dag}$ To whom correspondence should be addressed. Tel: +1 617 552 1332; Fax: +1 617 552 2011; Email: clote@bc.edu
}

\begin{document}
\maketitle

\begin{abstract}
Several algorithms for RNA inverse folding have been used to design
synthetic riboswitches, ribozymes and thermoswitches, whose activity
has been experimentally validated. 
The {\tt RNAiFold} software is unique
among approaches for inverse folding in that (exhaustive) constraint programming
is used instead of heuristic methods. For that reason, {\tt RNAiFold} 
can generate all sequences that fold into the target structure, or 
determine that there is no solution.
{\tt RNAiFold 2.0} is a complete overhaul of {\tt RNAiFold 1.0},
rewritten from the now defunct COMET language to C++. The new code
properly extends the capabilities of its predecessor by providing a
user-friendly pipeline to design synthetic constructs having the
functionality of given Rfam families. In addition, the new software
supports amino acid constraints, even for proteins translated in
different reading frames from overlapping coding sequences; moreover,
structure compatibility/incompatibility constraints have been
expanded. With these features, {\tt RNAiFold 2.0} allows the
user to design single RNA molecules as well as hybridization complexes of 
two RNA molecules.

\noindent
Availability: 
The web server, source code and linux binaries are publicly accessible at
\url{http://bioinformatics.bc.edu/clotelab/RNAiFold2.0}. 
\end{abstract}

\section{Introduction}

RNA inverse folding is the problem to determine one (or all) RNA sequences,
whose minimum free energy (MFE) secondary structure is identical to a
given {\em target} secondary structure. Most algorithms for inverse
folding use heuristics, such as ensemble defect optimization \cite{Zadeh.jcc11},
genetic algorithms \cite{Taneda.aabc11,frnakenstein,EsmailiTaheri.bb15}, 
simulated annealing \cite{Weinbrand.b13}, initial sequence optimization
\cite{Busch06}, adaptive walk \cite{ivoThesis,Gruber08}, etc.
Some algorithms, such as {\tt NUPACK-Design}, do not attempt to solve
the inverse folding problem, but instead minimize ensemble defect, which
measures the extent to which low energy structures 
deviate from the target structure.
In contrast to other approaches, {\tt RNAiFold} \cite{GarciaMartin.jbcb13}
is the only exhaustive, nonheuristic method, achieved by Constraint Programming 
(CP). {\tt RNAiFold} has been used to computationally design
functional, synthetic ribozymes, whereby cleavage kinetics have 
been experimentally determined \cite{Dotu.nar15}, 
and to detect novel IRES-like (internal ribosomal entry site) 
elements validated by a luciferase reporter assay \cite{Dotu.rb13}.
In this note, we describe differences between {\tt RNAiFold 1.0}
\cite{GarciaMartin.nar13} and {\tt RNAiFold 2.0}. 

First, {\tt RNAiFold 2.0} is a complete overhaul and reimplementation of all
the algorithms from \cite{GarciaMartin.jbcb13} in C++ using the new OR-Tools 
engine \url{https://code.google.com/p/or-tools/}.
Since the COMET engine for {\tt RNAiFold 1.0} is now obsolete, 
and no new licenses will be issued, it is not possible for users 
to execute our COMET source code. In contrast, our new code is now 
available along with the publicly available engine OR-Tools, supported by
Google. Both can be installed and executed by users on various
Operating Systems with a C++ compiler.
Second, {\tt RNAiFold 2.0} allows the user to require solutions to
be {\em compatible} with a second given structure, in addition to folding
into the target structure, and/or be {\em incompatible} with base pair
formation at positions listed in a {\em prohibition} list. Moreover, 
amino acid constraints have been added, requiring solutions not only to 
fold into a target structure, but also to code a given protein 
(or to code for the most similar protein, as determined by the 
BLOSUM62 similarity matrix). In addition, the user can choose to use
Turner'99 [resp. Turner'04] energy parameters  \cite{Turner.nar10}
by interfacing with Vienna RNA Package 1.8.5 or 2.1.7 \cite{Lorenz.amb11}.
Third, {\tt RNAiFold 2.0} provides a distinct, novel web service for 
a fully automated pipeline to design synthetic RNAs, such as the 
synthetic hammerheads described in \cite{Dotu.nar15}.
In this case, the user can specify a family from 
Rfam 12.0 \cite{Nawrocki.nar14}, then select a member 
of the automatically generated list of Rfam seed sequences whose 
minimum free energy (MFE) structure coincides with the (functional) 
Rfam consensus structure, and then set a threshold for sequence conservation. 
{\tt RNAiFold 2.0} then computes a list of synthetic RNAs, 
which fold into the (functional) target Rfam consensus structure, 
and are guaranteed to contain those presumably important nucleotides 
located at positions which exceed the user-specified sequence 
conservation threshold.

The plan of this paper is to discuss (1) the automated synthetic
design pipeline, (2) amino acid sequence and prohibited base pair
constraints -- all of which are not present in our earlier software
{\tt RNAiFold 1.0} \cite{GarciaMartin.nar13} -- and to present 
(3) a comparison of
{\tt RNAiFold 2.0}, {\tt RNAiFold 1.0}, and other inverse
folding software.

\enlargethispage{-65.1pt}

\section{Rfam-base design pipeline}

Details of the novel method for synthetic RNA design are described in
Dotu et al., \cite{Dotu.nar15}, which additionally discusses 
the selection criteria (pointwise entropy, ensemble defect, etc.)
used to prioritize synthetic type III hammerhead 
ribozyme candidates for experimental validation.  The
{\tt Rfam-based design} pipeline is now an integral part of the 
{\tt RNAiFold 2.0} web server, so we describe here
how to fill in the web pages displayed in 
Figures~1 and 2,
in order to design synthetic RNAs likely to
function similarly to RNAs in a user-specified family from 
Rfam 12.0 \cite{Nawrocki.nar14}. See
the on-line manual for more information.

{\bf Step 1, depicted in the left panel of
Figure~1:}
Though not required, it is useful to enter an email address for 
notification when the computation terminated in the case of a long job. 
First, the user should decide whether to check the checkbox that 
subsequently allows the the selection of sequences whose MFE structure is
perhaps not {\em identical} to the Rfam consensus structure (explained below).
Next, select an Rfam family -- in the case of hammerhead III ribozymes, 
this is RF00008. Next, select the energy model -- either Turner'99 or 
Turner'04, for which energy parameters are described in \cite{Turner.nar10}.
It is commonly held that Turner'04 parameters are more accurate, though
this is not necessarily the case, since 
Vienna RNA Package {\tt RNAfold} \cite{Lorenz.amb11}  predicts the 
correct, functional structure for Peach Latent Mosaic Viroid (PLMVd) 
hammerhead ribozyme AJ005312.1/282-335
using the Turner'99 parameters 
(left panel of Figure~1, left image), 
while the incorrect structure is 
predicted using the Turner'04 parameters
(left panel of Figure~1, right image). 
Choose the treatment of dangles (stacked, single-stranded nucleotides), where
choices are no dangle (-d 0), the minimum of $5'$- and $3'$-dangle (-d 1),
the sum of $5'$- and $3'$-dangle (-d 2), 
minimum of $5'$- and $3'$-dangle plus coaxial stacking (-d 3).
For design of functional hammerheads in \cite{Dotu.nar15},
we used the Turner'99 model with (-d 1), since the minimum free energy
structure of PLMVd AJ005312.1/282-335 is identical to the Rfam
consensus structure.

\begin{figure*}
\begin{subfigure}[b]{0.46\textwidth}
\includegraphics[width=\textwidth]{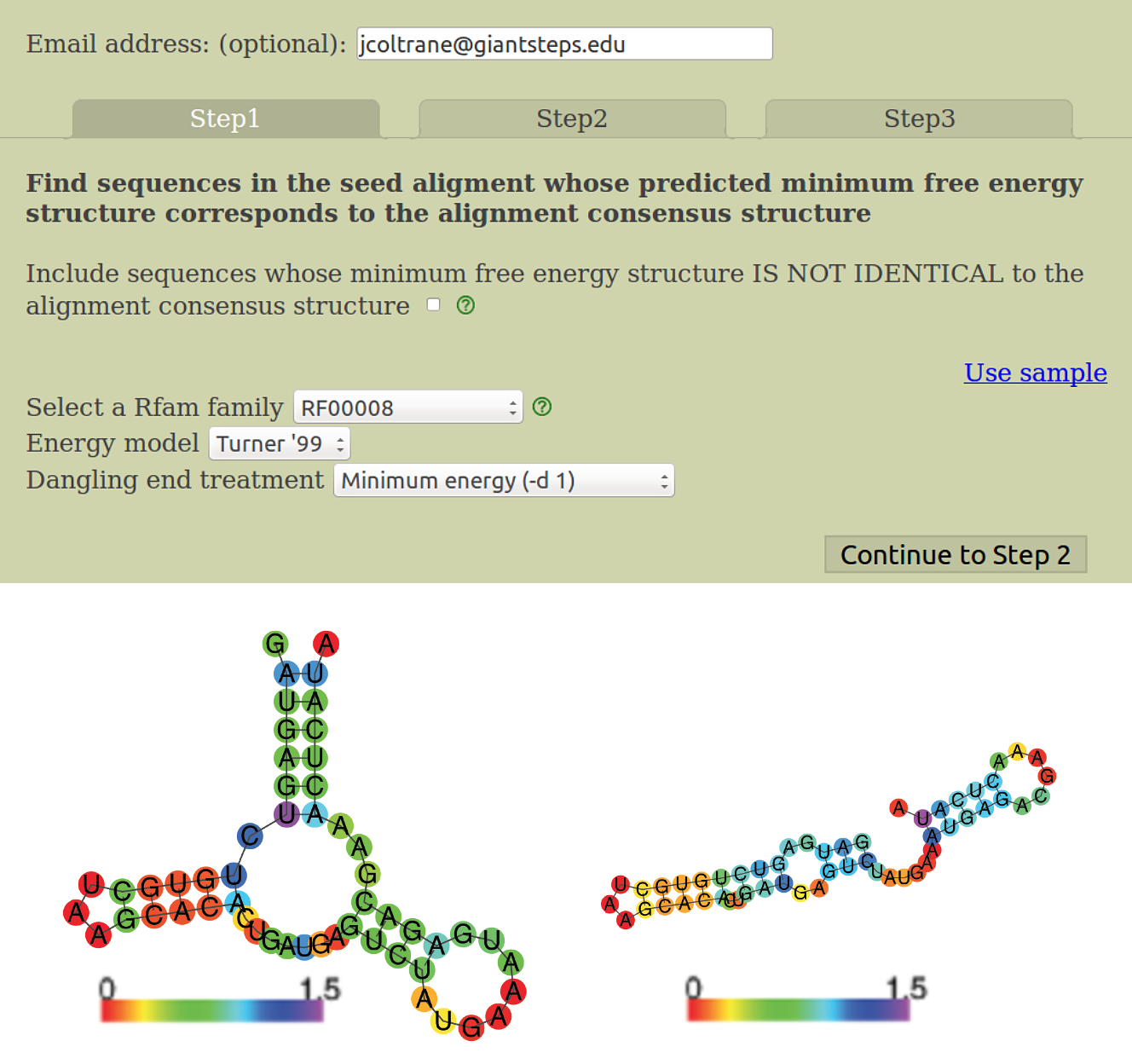}
\caption{}
\label{fig:A1}
\end{subfigure}%
\qquad
\begin{subfigure}[b]{0.46\textwidth}
\includegraphics[width=\textwidth]{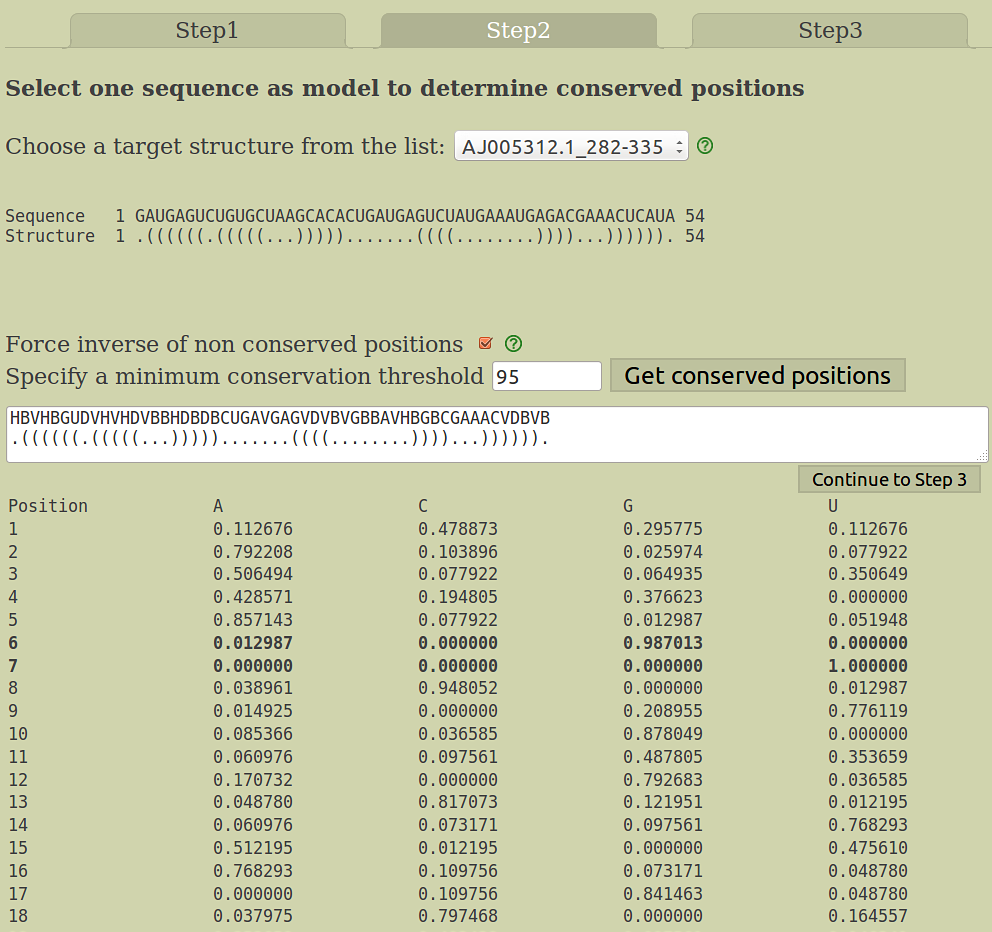}
\caption{}
\label{fig:B1}
\end{subfigure}
\caption{
{\em (a)}
First of 3 screens in {\em RNA Rfam-based Design}, invoked to
automate the generation of synthetic RNAs likely to adopt the same function
as RNAs in a user-specified Rfam class. Computations may take long, so
though optional, it is advisable to enter an email address to be informed
of the results when ready. The user must select an Rfam family and
the energy model, i.e.
Turner'99 or Turner'04 \cite{Turner.nar10} together with a dangle state.
As shown in the figure, the Turner'99 parameters (Vienna 1.8.5) can
prove to be a better choice than the Turner'04 parameters (Vienna 2.1.7)
in certain circumstances -- here, Vienna 1.8.5 predicts the correct,
functional structure for the hammerhead type III ribozyme (left image) from
Peach Latent Mosaic Viroid (PLMVd) with accession code AJ005312.1/282-335,
while the Vienna 2.1.7 predicts the incorrect structure (right image).
{\em (b)}
Second of 3 screens in {\em RNA Rfam-based Design}, where the user
selects a sequence in the pull-down menu; this sequence, which belongs to
the chosen Rfam family will serve as an initial {\em model} to generate
synthetic sequences.  Each displayed sequence folds into
the Rfam consensus structure when using the selected energy parameters
(if no sequence is shown, then no sequence has this property).
In this screen, the
user may specify that {\tt RNAiFold 2.0} automatically generate sequence
constraints for positions that are conserved in the Rfam seed alignment
to user-specified minimum threshold; to avoid generating solutions that
are too similar to the model sequence, the server automatically
generates IUPAC constraints to {\em disagree} with the model sequence
at all positions where the seed alignment has less than the specified
conservation threshold. The position-specific compositional frequency
(profile) of the Rfam seed alignment is displayed for each position.
}
\label{fig:screenShotDesign1-2}
\end{figure*}

{\bf Step 2, depicted in the right panel of
Figure~\ref{fig:screenShotDesign1-2}:}
A pull-down menu allows one to select a target structure
from those Rfam seed alignment sequences, whose
MFE structure is identical (or similar)
to the corresponding Rfam consensus stucture. 
The Rfam consensus structure for a given sequence is determined by
placing base pairs in positions dictated by the Rfam consensus (indicated
by angle brackets at the bottom of alignments in Stockholm format), then
removing base pairs if the nucleotides do not form a Watson-Crick or wobble
pair, and finally removing base pairs at positions $i,j$ when $i<j<i+4$.
If the user did not check the checkbox which allows consideration of
sequences, whose MFE structure is not identical with the  Rfam
consensus structure, then it can happen that no target structure will 
be displayed -- indeed, this will happen if
no Rfam sequences fold using Turner parameters into their corresponding
consensus structure. 
If the checkbox was checked in Step 1, then sequences
whose MFE structure closely resembles the Rfam consensus structure will be
displayed, and the base pair distance between the consensus and MFE structure
will be indicated in parenthesis.
After selecting a target structure, the user should set a conservation
threshold $\theta$, whose default value is 95\%. The server determines
the compositional frequency as each position of the selected structure,
and sets a {\em sequence constraint} for those positions whose compositional
frequency exceeds threshold $\theta$. The user may check the box which
additionally sets a sequence constraint for all remaining positions to
be different from the nucleotide of the Rfam sequence whose target struture
has been selected -- for instance, the largest nucleotide frequency at
position 1 of the Rfam alignment is $P(C)= 0.478873$, which is less than
the conservation threshold of $0.95$, and since Rfam sequence
AJ005312.1/282-335 contains G at position 1, the sequence constraint
contains IUPAC code H (not G) at position 1. The resulting target
structure and sequence constraint is then displayed.

{\bf Step 3, depicted in the left panel of
Figure~2:}
By clicking on the button {\em Continue to Step 3}, located in the
bottom right of Step 2, the FASTA comment, target structure,
sequence constraints, energy model, dangle treatment, etc. are 
automatically entered in the appropriate places in the form in Step 3.
The user can choose to generate 1, 5, 10, 50, or MAX (maximum number
of solutions that can be computed within a system-dependent
resource bound on computation time). Additional constraints can be
added at this point. For instance, the user could require all
solutions be compatible with the additional structure
{\tiny \tt ................(((((.......))))).....................}
for which {\tt RNAiFold 2.0} correctly returns the solution
AGGCGGUAAC
CCGAUCCGGG
UCUGAAGAGC
UCGAGUUAAA
GGGCGAAACC
GCCC.
\begin{quote}
\begin{tiny}
\bf
\begin{verbatim}
> Example 1: Designing synthetic hammerhead ribozymes
.((((((.(((((...))))).......((((........))))...)))))).
HBVHBGUDVHVHDVBBHDBDBCUGAVGAGVDVBVGBBAVHBGBCGAAACVDBVB
#MAXsol
1
#dangles
1
#RNAcompstr
................(((((.......))))).....................
#temp
37
#LNS
0
\end{verbatim}
\end{tiny}
\end{quote}
Note that (exhaustive) Constraint Programming (CP) is used,
rather than (heuristic) Large Neighborhood Search (LNS),
as indicated by 0 following \#LNS -- this is the default, unless
otherwise indicated. 
Additional constraints can be included in the input file, by using the 
appropriate label preceeded by the ``pound'' symbol (``\#''),
where the desired value appears in the next line (see on-line manual
for more details). 
When running a local copy of the executable,
one uses command-line flags, as in
{\tt RNAiFold2\_2.1.7 -RNAscdstr '(((...)))' -RNAseqcon NNNAAANNN}.

The right panel of Figure~2 shows the
output of the {\tt RNAiFold 2.0} web server when designing
hammerhead ribozymes -- cf. ``Use sample'' link visible in 
left panel of Figure~1.
A pull-down menu displays each of the solutions found within the
system-dependent time limit.
For each solution, the secondary structure is displayed
(identical to the target structure) and as well as summary information 
for GC-content, Boltzmann probability of target MFE structure, average
pointwise entropy (also called positional entropy) \cite{Huynen.jmb97},
Morgan-Higgs and Vienna structural Diversity \cite{morganHiggsBarrier},
expected base pair distance from target \cite{GarciaMartin.jbcb13}, 
and ensemble defect \cite{RNAdesignNAR2004}. These measures can be
used to prioritize the selection of candidates for experimental validation
\cite{Dotu.nar15}. Finally, a link to NCBI {\tt BLAST} is provided to
search for sequences similar to the 
{\tt RNAiFold 2.0} solution sequences.

\section{Novel features in RNAiFold 2.0}

Apart from the synthetic RNA design pipeline described in the
previous section, {\tt RNAiFold 2.0} provides a number of novel features 
not available in {\tt RNAiFold 1.0}:
(a) the user can choose to use either the Turner’99 or Turner’04
energy parameters 
by the built-in interface with
Vienna RNA Package 1.8.5 or 2.1.7;
(b) the target can be specified using expanded dot-bracket notation,
where a comma indicates that the position may be paired or not; i.e.
{\tt RNAiFold 2.0} now supports {\em partial targets};
(c) structural constraints have been expanded -- in addition to folding into
the target structure, solutions can be required to be {\em compatible}
with an additional structure, and can be required to be {\em incompatible}
with base pair formation at those positions listed in a {\em prohibition list};
(d) amino acid constraints have been added, which require solutions 
not only to fold into a target structure, but also to code a given protein
(or to code for the most similar protein, as determined by
the BLOSUM62 similarity matrix);
(e) the flag {\tt RandomAssignment} can be set, which randomizes 
instantiation order of variable values
(used to provide a more unbiased sample of solutions when search space is 
very large).

Base pair formation may be prohibited using 3 different syntaxes.
(i) If a secondary structure $\seq$ is listed after the flag 
{\tt IncompBP}, then
positions $(i,j)$ where a base pair occurs in $\seq$ are not allowed to pair
in every solution returned. 
(ii) The syntax {\bf P i j k} may be used, which prevents position $i$ from
pairing with $j,j+1,j+2,\ldots,j+(k-1)$.
(iii) A comma separated list of pairs $i_1 j_1$,\ldots, $i_n j_n$
can be specified, which prevents position $i_1$ from pairing with $j_1$,
position $i_2$ from pairing with $j_2$, etc. The user may combine syntax
from (ii) and (iii) together, as shown below. Since
structural compatibility constraints were illustrated in 
the previous section, we illustrate the use of constraints
(a)-(d) without again demonstrating structural compability constraints.

The user first selects the energy model (Turner'99
or Turner'04) from the web page -- here, Turner'04 was selected. 
Additional parameters can be set within
the web page form, or within a command file that is uploaded, and shown below. 
Defaults are taken, unless otherwise mentioned in the command file or
web page. Consider the following example,
where a stem-loop partial structure is given on the left, with A at
position 1, a GNRA-tetraloop at positions 6-9, and a partial structure
consisting of a base pair $(17, 26)$ as indicated on the right fragment
of the structure.  Commas appear at position 16, 18-25, 27, to specify
that these positions may be paired or unpaired. Additionally,
the first position is prohibited from pairing with any other nucleotide in this
27-nt sequence (P 1 2 26); position 3 is prohibited from pairing with 
positions 16 and 17 (P 3 16 2), 
and the nucleotides at positions 4,17 and 4,18 and
4,19 are prohibited from pairing.
\begin{quote}
\begin{tiny}
\bf
\begin{verbatim}
> Example 2: Partial target, IUPAC codes, incompatible base pairs
.((((....))))..,(,,,,,,,,),
ANNNNGNRANNNNNNNNNNNNNNNNNN
#MAXsol
1
#IncompBP
P 1 2 26,P 3 16 2,4 17,4 18,4 19
#dangles
1
#temp
37
#LNS
0
\end{verbatim}
\end{tiny}
\end{quote}
The solution returned in 2.18 seconds is the following:
\begin{quote}
\begin{tiny}
\bf
\begin{verbatim}
.((((....))))..,(,,,,,,,,),
Init time: 0
AGGGGAAAACCCCAAGGAGCAAAGCCC
GC content: 0.59 - AUs: 0 - GCs: 5 - GUs: 0
.((((....))))..((.((...))))
Probability of MFE structure:0.612535
Expected pointwise entropy:0.298331
Morgan-Higgs structural Diversity:4.41067
Vienna structural Diversity:2.98779
Expected base pair distance:3.29118
Ensemble defect:5.8756
Search time: 2.18
Time elapsed 2.18
\end{verbatim}
\end{tiny}
\end{quote}
The first structure in the output is the user-specified {\em partial} target.
The solution is given in the third line, followed by GC-content, number of
base pairs of each type, and the MFE structure of the solution. As required,
the MFE structure of each solution agrees with the target (partial) structure 
at positions occupied by a dot, left parenthesis,
or right parenthesis, but may differ in positions corresponding to
commas in the target (partial) structure. 
No U's occur in positions 2-27, as required by the prohibition P 1 2 26,
nor can 3 form a base pair with 16,17, nor can 4 form a base pair with
17,18,19. 
Finally, 
note that the command file may be uploaded, or its contents may be copied
into the web form text area when the option ``paste input'' is selected.
This may save time, if many options and parameters need to be given.

Finally, amino acid constraints may be specified by using the flag
{\tt AAseqcon}, followed by one or more amino acid sequences, followed by
the flag {\tt AAstartPos} after which the starting position of the first
codon of each amino acid sequence is given. Note that there is no bound on the
number of (possibly overlapping) coding regions for distinct peptides.
In the following example,
the target structure has length 52, positions 1-51 code for the peptide
FFREDLAFPQGKAREFS and positions 2-52 code for the peptide FLGKIWPSHKGRPGNFL.
The flag {\tt AAsimilCstr} specifies whether the solution must 
exactly code the given peptides (value 5), or whether each amino acid
coded by the returned solutions must have BLOSUM62 similarity at least
$x$ (for value $x<5$) with each amino acid of the given peptide. Values
6,7 allow the user to enter specific symbols that designate chemical properties
of residues coded by all solutions returned -- e.g. hydrophobic, 
positively or negatively charged, polar, etc. (see on-line manual for details).
\begin{quote}
\begin{tiny}
\bf
\begin{verbatim}
> Example 3: overlapping amino acid constraints
......((((((..((((((((((((....))))))))))))...)))))).
UUUUUUANNNNNNNNNNNNNNNNNNNNNNNNNNNNNNNNNNNNNNNNNNNNN 
#AAseqcon 
FFREDLAFPQGKAREFS,FLGKIWPSHKGRPGNFL
#AAstartPos 
1,2 
#MAXsol 
0 
#AAsimilCstr 
3
\end{verbatim}
\end{tiny}
\end{quote}
Note that flag {\tt MAXsol} has the value 0, which allows the user
to run {\tt RNAiFold 2.0} locally with no upper bound on the number of
solutions returned. In this case, {\tt RNAiFold 2.0} will either teminate
with all possible solutions, or the process will die after memory
exhaustion, or the user can terminate the process; however, in all cases
the output can be saved to a file.

Another novel feature of amino acid constraints is that the flag
{\tt MaxBlosumScore} allows {\tt RNAiFold 2.0} to determine a solution
of inverse folding for which the BLOSUM62 similarity to the target
peptide is an absolute maximum; i.e. no other solution of inverse folding
codes a peptide having larger BLOSUM62 similarity to the specified target
peptide.
\section{Comparison with other software}

Tables~1,2 and Figure~3 provide a comparison of all current inverse folding 
software and web servers. Table~1 gives an overview of distinctive features
of each software, while Table~2 gives an overview of the quality and 
quantity of solution sequences returned by each method. 

To our knowledge, the only other web servers or software for RNA design are:
{\tt ERD} \cite{EsmailiTaheri.bb15},
{\tt FRNAkenstein} \cite{frnakenstein},
{\tt Incarnation} \cite{Reinharz.b13}, 
{\tt Info-RNA} \cite{Busch06},
{\tt MODENA} \cite{Taneda.aabc11},
{\tt NUPACK} \cite{Zadeh.jcc11},
{\tt RNA-SSD} (web server is called RNAdesigner)
\cite{Andronescu04},
{\tt RNAfbinv}  \cite{Weinbrand.b13},
{\tt RNAinverse} \cite{Lorenz.amb11}.
Default parameters were used for all software, with the exception of
{\tt RNAinverse}, where we used flags {\tt -R -1}.
In addition to the
features indicated in Table~1, {\tt RNAiFold} (both versions
1.0 and 2.0) is to our knowledge the only software that allows
the user to stipulate exact GC-content range for the sequences returned;
although {\tt Incarnation} is claimed to support GC-content range stipulation,
it returns some sequences that violate the user-stipulated GC-content.
{\tt RNAdesign} allows the user to designate
a ``target'' GC-content; however, this only introduces a bias 
trying to adjust the GC-content of the output sequences.

{\tt RNAiFold 2.0} also allows the user to stipulate the energy range 
$E(\aseq,S_0)$ for the sequences $\aseq$ returned, where $S_0$ denotes the
target structure.
The {\tt ERD} web server (http://mostafa.ut.ac.ir/corna/erd-cons/) claims to
support energy range stipulation, but in a test
only one of the 10 requested solutions
for target structure {\tt ((((...)))).} had energy in the requested
range of -4 to -2 kcal/mol.
{\tt RNAiFold 2.0} can as well return that sequence which folds into the target
$S_0$ and has minimum energy among all such sequences; however, this feature is
not a true constraint and results from the fact that {\tt RNAiFold 2.0}
generates all solutions. Apart from other features indicated in Table~1,
{\tt NUPACK} is the only software that solves the inverse
folding problem for both DNA and RNA, and in the case of DNA, allows the
user to stipulate magnesium and sodium ion concentrations.

To compare {\tt RNAiFold 2.0} with {\tt RNAiFold 1.0} and other existent
inverse folding software, we used 63 target structures,
ranging in size from 54 to 1398 and used in earlier
benchmarking from \cite{GarciaMartin.nar13} (data described later). 
For the comparison of versions 1.0 and 2.0 of {\tt RNAiFold}, 
we ran each program 
100 times per target structure, for each of the 63 target structures 
just described, Each run had a time upper bound of 10 minutes; 
however, execution was terminated as soon as the first solution was returned.
A solution was returned on average within approximately 10 seconds
(10.78 s for {\tt RNAiFold 2.0}, 12.64 seconds for {\tt RNAiFold 1.0}).
The web server {\tt RNAiFold 2.0} obtained 
more solutions than {\tt RNAiFold 1.0} for $\sim 10$\% of the data; i.e.
24 targets had more solutions, 21 targets had the same number, and
18 targets had fewer solutions.  {\tt RNAiFold 2.0} returned solutions
in less time than {\tt RNAiFold 1.0} for $\sim 43$\% of the data; i.e.
38 targets were solved more quickly with version 2.0,
14 targets required the same time (or neither version returned a solution
within 10 minutes),
11 targets were solved more quickly with version 1.0.
Benchmarking data and target structures 
can be found on web server at the tab `Download'.

Table~2 illustrates differences in the quantity and quality of 
sequences returned for a given target structure. Measures that
quantify the extent to which the ensemble of low energy structures of
a given sequence resembles a target structure (ensemble defect, expected
base pair distance) or how {\em diverse} structures are from each other
(Morgan-Higgs and Vienna structural diversity) are defined in the next
paragraph.  For each of the
63 target structures, each software was run 10 minutes to generate a
quantity of sequences using default settings.  {\tt ERD} returns an output
100\% of the time, where 85\%  of the output sequences fold into the
target structure. In contrast, {\tt RNAiFold 2.0} returns an
output 65\% of the time, but 100\% of its output is guaranteed to fold
into the target structure. Incarnation returns 41,535 sequences on average
for each target, but less than 0.2\%  fold into the target structure,
while {\tt RNAiFold 2.0} returns 55,476 sequences on average and
100\% fold into the target structure. {\tt Info-RNA} has over 72\% GC-content,
due to the initial choice of starting sequence, while 
{\tt NUPACK} and {\tt RNAiFold 2.0} have around 57\% GC-content (and moreover,
{\tt RNAiFold 2.0} allows the user to set a desired GC-content range), 
while {\tt RNA-SSD} has close to 36\% GC-content.

{\em Benchmarking Data:}
All benchmarking data is available at
\url{http://bioinformatics.bc.edu/clotelab/RNAiFold2.0} (tab `Download'). 
It is comprised of
dataset A from \cite{Taneda.aabc11} and datasets B,C from
\cite{Andronescu04}.
Dataset A consists of 29 target structures first described in
\cite{Taneda.aabc11}, obtained by taking the Rfam structure of the
longest sequence from the seed alignment for each of
families RF00001-RF00030 in the Rfam 9.0 database \cite{Gardner.nar09},
with the exception of family RF00023 (tmRNA).
Dataset B consists of 24 target structures, whose GenBank accession codes
are given and which were first described in
\cite{Andronescu04}; dataset C consists of 10 longer target structures
described as follows -- see \cite{Andronescu04} for references.
{(1)}
minimal catalytic domain of the hairpin ribozyme satellite
RNA from tobacco ringspot virus.
{(2)}
U3 snoRNA $5'$-domain from {\em Chlamydomonas reinhardtii}.
{(3)}
5S rRNA from {\em H. marismortui}.
{(4)}
VS Ribozyme from {\em Neurospora mitochondria}.
{(5)}
R180 ribozyme.
{(6)}
XS1 ribozyme, {\em Bacillus subtilis} P RNA-based ribozyme.
{(7)}
RNase P RNA from {\em H. sapiens}.
{(8)}
S20 mRNA from {\em E. coli}.
{(9)}
RNAse P RNA from {\em Halobacterium cutirubrum}.
{(10)}
Domains 1,3,5 from the group II intron ai5$\gamma$ from the mitochondria of
{\em S. cerevisiae}.
(References for experimentally determined structures not given due to
space constraints; see \cite{Busch06} for these references.) 


\medskip

\noindent
{\em RNA structural measures:}
A secondary structure $S$ may be considered as the set of its base pairs,
hence $(i,j)$ is a base pair of $S$ when $(i,j) \in S$, and the collection
of base pairs common to structures $S,T$ is denoted $S \cap T$.
Many RNA structural measures can be defined from the base pairing probabilities,
computed by McCaskill's algorithm \cite{mcCaskill} and implemented in
{\tt RNAfold -p} \cite{hofacker:ViennaWebServer,Lorenz.amb11}.  
Given the RNA sequence $\aseq = a_1,\ldots,a_n$, let
$p_{i,j} = \sum_{\{ S: (i,j) \in S\}} P(\aseq,S) =
\sum_{\{ S: (i,j) \in S\}} \exp(-E(\aseq,S)/RT)/Z$,
where $P(\aseq,S)$ is the Boltzmann probability of structure $S$ of 
RNA sequence  $\aseq$,
$E(\aseq,S)$ is the Turner energy of secondary 
structure $S$ \cite{turner,xia:RNA},
$R \approx 0.001987$ kcal mol$^{-1}$ K$^{-1}$ is the universal gas 
constant, $T = 310.15$ is
absolute temperature, and the {\em partition function}
$Z = \sum_{S} \exp(-E(\aseq,S)/RT)$, where the sum is taken over all
secondary structures $S$ of $\aseq$. Symmetrize the base pair probabilities,
by defining $p_{j,i}=p_{i,j}$ for $1 \leq i,j \leq n$, $i \ne j$, 
and define $p_{i,i} = 1- \sum_{i \ne j} p_{i,j}$ to be the probability
that position $i$ is unpaired. Let $s_0$ denote the minimum free energy
structure of input RNA sequence $\aseq$.
\medskip

\noindent
{(1)}
{\em Morgan-Higgs diversity} \cite{morganHiggsBarrier}
is defined for $\aseq = a_1,\ldots,a_n$ by
$n - \sum_{S,T}  \sum_{j=1}^n
P(S) \cdot P(T) \cdot \sum_{i=1}^n I[ (i,j) \in S \cap T \mbox{ or }
(j,i) \in S \cap T]$,
where the sum is taken over all secondary structures $S,T$ of $\aseq$,
and $I$ denotes the indicator function.
\smallskip

\noindent
{(2)}
{\em Vienna diversity} is defined for $\aseq = a_1,\ldots,a_n$ by
$\sum_{1 \leq i<j \leq n}
p_{i,j}(1-p_{i,j}) +
(1-p_{i,j})p_{i,j} =
\sum_{i<j}
2p_{i,j}(1-p_{i,j})$.
In \cite{Gruber.bb08}, this is called {\em ensemble diversity}.
\smallskip

\noindent
{(3)}
{\em Expected number of base pairs} is defined by
$\sum_{1 \leq i<j \leq n} p_{i,j}$.
\smallskip

\noindent
{(4)}
{\em Expected base pair distance} \cite{GarciaMartin.jbcb13}
is defined to the MFE structure
$s_0$ of input RNA sequence $\aseq$ is defined by
$\sum_{1 \leq i<j \leq n}
I[ (i,j) \not\in s_0 ] \cdot p_{i,j} +
I[ (i,j) \in s_0 ] \cdot (1-p_{i,j}),
$
where $I$ denotes the indicator function. 
\smallskip

\noindent
{(5)}
{\em Ensemble defect} \cite{RNAdesignNAR2004}
is the expected number of nucleotides whose base pairing status differs
from the target structure $S_0$,
defined by
$n - \sum_{i \ne j} p_{i,j} \cdot I[ (i,j) \in S_0 ]
- \sum_{1 \leq i \leq n} p_{i,i}\cdot I[ \mbox{$i$ unpaired in $s_0$}]$,
where $I$ denotes the indicator function.  
\smallskip

\noindent
{(6)}
Given sequence $\aseq = a_1,\ldots,a_n$, the
{\em total positional entropy} (H) \cite{Huynen.jmb97} is defined by
$H(\aseq) = \sum_{i=1}^n 
\left\{  - \left( p_{i,i} \cdot \ln p_{i,i} + (1- p_{i,i}) \cdot 
\ln (1-p_{i,i}) 
\right) \right\}$, where $0 \cdot \ln 0$ is defined to be $0$.
{\em Normalized positional entropy} for $\aseq$ is $H(\aseq)/n$; the
average normalized positional entropy is the average normalized positional
entropy, where the average is taken over all output sequences.
\smallskip

\noindent
{(7)}
Given the RNA sequence $\aseq = a_1,\ldots,a_n$ and target structure $S_0$,
the {\em expected proportion of native (i.e. target) contacts} 
$\sum_{(i,j) \in S_0} p_{i,j}/|S_0|$, 
where $|S_0|$ denotes the number
$\sum_{(i,j) \in S_0} p_{i,j}/|S_0|$, where $|S_0|$ denotes the number
of base pairs in $S_0$.  The average is then taken over all sequences
output for a given target structure, and then the average is taken over all
targets. This value is called ``mean fraction of bases retained'' in
\cite{Pei.bg15}, where it is approximated by sampling using {\tt RNAsubopt}
\cite{wuchtyFontanaHofackerSchuster,Lorenz.amb11}.

\section{Results}

In this section, we illustrate how to use {\tt RNAiFold 2.0}, in order to
design synthetic RNAs that trigger two biologically significant {\em recoding} 
events: (1) SECIS elements cause the ribosome to incorporate
a non-standard amino acid, {\em selenocysteine}, into a growing peptide chain;
(2) programmed -1 ribosomal frameshift signals cause the ribosomal reading 
frame in messenger RNA to shift at a specific site by -1 within the coding 
region. 

Prokaryotes, archaea, and eukaryotes employ the UGA stop codon to code for
selenocysteine, rather than terminating protein translation, provided that
a {\em selenocysteine insertion} (SECIS) element occurs downstream of the
UGA stop codon. The SECIS element is a $\sim 42$ nt sequence having conserved
nucleotides at certain positions, which folds into a stem-loop 
secondary structure \cite{commansBoeck} -- see target structure in Example 5.
In prokaryotes, the SECIS element lies immediately after the UGA stop 
codon, while in eukaryotes and archaea it lies in the $3'$ untranslated region
\cite{GrundnerCulemann.r99}.
In the formate dehydrogenase F (fdhF) gene of
{\em Salmonella enterica} 
(GenBank: CDS70432.2), the 42-nt sequence
UGACACGGCC CAUCGGUUGC AGGUCUGCAC CAAUCGGUCG GU
consists of the UGA stop codon immediately followed by
the SECIS element. This sequence folds into the stem-loop structure
shown in Example 6, and codes the 14 residue peptide
UHGPSVAGLHQSVG (`U' denotes selenocysteine).

In contrast, the homologous 14 residue peptide of the fdhF protein
of {\em Raoultella ornithinolytica} is given by CHGPSVAGLQQALG, where
cysteine appears instead of selenocysteine. 
Unlike {\em S. enterica}, the $42=14 \cdot 3$ nt
portion of the fdhF gene of {\em R. ornithinolytica} (Genbank AJF73661.1) 
begins with UGC, which codes for cysteine, rather than UGA, a stop codon
which codes for selenocysteine in the presence of a SECIS element; 
moreover, the 42-nt sequence
of {\em R. ornithinolytica} does not fold into a stem-loop SECIS structure.

The following input file defines the target structure to be the
MFE structure of the 42-nt RNA from {\em S. enterica}, sets as sequence
constraints the bulged U18 and GGUC hairpin identity (known to be important 
for SECIS functionality \cite{Liu.nar98,Sandman.nar03}, and
sets as amino acid constraints the 14-mer of
{\em R. ornithinolytica}, with `C' replaced by `U'.
\begin{quote}
\begin{tiny}
\bf
\begin{verbatim}
> Example 6: Selenocystein insertion in AJF73661.1 140
.....(((((.((.(((.(((....)))))).)).)))))..
NNNNNNNNNNNNNNNNNUNNNGGUCNNNNNNNNNNNNNNNNN
#MAXsol
0
#AAtarget
UHGPSVAGLQQALG
#AAsimilCstr
-1
#MaxBlosumScore
1
#dangles
2
#LNS
1
\end{verbatim}
\end{tiny}
\end{quote}
In $0.24$ seconds {\tt RNAiFold 2.0} determined the optimal solution 
UGACACGGGC CCUCGCUUGC AGGUCUGCAG CAAGCGCUCG
GA, which begins by the UGA stop codon, translates the 14-mer 
UHGPSVAGLQQALG, and folds into the requisite
SECIS stem-loop. This example shows how {\tt RNAiFold 2.0} can be used
to re-engineer selenoproteins from cysteine-bearing proteins.

In the retrovirus HIV-1, Pol is obtained from a fused
Gag-Pol polyprotein via a programmed -1 ribosomal frameshift, which
is caused by a heptameric {\em slippery
sequence} (U UUU UUA), where the Gag reading frame is indicated, 
together with a downstream {\em frameshift stimulating stem-loop} structure
\cite{Ofori.jmc14}. 
Using the target and constraints from Example 2, we ran {\tt RNAiFold 2.0} to
find the complete set of 29,340 solutions in 539.49 seconds ($\approx 9$ hours).
The sequences returned by {\tt RNAiFold 2.0} constitute synthetic putative
ribosomal frameshift signals, which could be tested for frameshift efficiency.
Additionally, we can infer the relative importance of amino acid 
coding requirements for Gag and Pol versus secondary structure requirements
within the frameshift signal,
by comparing naturally occurring -1 ribosomal frameshift elements in
Rfam family RF00480 with the solutions returned by {\tt RNAiFold 2.0}.

\section{Conclusion}

{\tt RNAiFold 2.0} is a complete overhaul and reimplementation of 
the algorithms from \cite{GarciaMartin.jbcb13} in C++ using the new OR-Tools
engine \url{https://code.google.com/p/or-tools/}. Novel features of the 
new software and web server, beyond those of
{\tt RNAiFold 1.0}, include an automated pipeline for synthetic RNA design,
use of Turner'99 or Turner'04 energy model, stipulation of a
partial target structure, stipulation of
prohibited (incompatible) base pairs, and amino acid constraints.
Given a target non-pseudoknotted hybridization complex of two structures,
{\tt RNAiFold 2.0} can output {\em pairs} of sequences, whose minimum
free energy hybridization complex is equal to the target. All the previously
described constraints are supported for hybridization -- see the on-line
manual section on {\em Cofold} for the syntax and an example.
Availability of the source code will allow users to design synthetic RNAs,
following the pipeline we used to design functional synthetic hammerhead 
ribozymes in \cite{Dotu.nar15}.

%

\section{FUNDING}

This work was supported by the National Science Foundation [DBI-1262439].

\subsubsection{Conflict of interest statement.} None declared.

\section{ACKNOWLEDGEMENTS}

This material is based upon work supported by the National Science 
Foundation under Grant No. DBI-1262439. We would like to thank anonymous 
reviewers for useful suggestions.

\hfill\break\newpage

\bibliographystyle{plain}
\bibliography{biblio}

\hfill\break\newpage

\begin{table*}[tbph]
\footnotesize
\begin{tabular}{|c|c|c|c|c|c|c|c|c|c|c|c|c|c|c|}
\hline
Software        	&    	$\Downarrow$	&    	 WS 	&    	PK 	&    	 H	&    	MT	&    	PT	&    	   T	&    	 EM	&    	 D	&    	SeqC	&    	 StrC	&    	AaC   	&    	O	&    	 Num	\\
\hline
{\tt RNAiFold 2.0}        	&    	\checkmark	&    	\checkmark	&    	\textemdash	&    	\checkmark	&    	\textemdash	&    	\checkmark	&    	\checkmark	&    	'99,'04	&    	0,1,2,3	&    	\checkmark	&    	\checkmark	&    	\checkmark	&    	mfe	&    	MAX	\\
{\tt RNAinverse} 	&    	\checkmark	&    	\checkmark	&    	\textemdash	&    	\textemdash	&    	\textemdash	&    	\textemdash	&    	\checkmark	&    	'99,'04	&    	0,1,2,3	&    	IUPAC$\star$	&    	\textemdash	&    	\textemdash	&    	mfe, prob	&    	100	\\
{\tt RNA-SSD} 	&    	\textemdash	&    	\checkmark	&    	\textemdash	&    	\textemdash	&    	\textemdash	&    	\textemdash	&    	\checkmark	&    	'99	&    	1	&    	IUPAC$\star$	&    	\textemdash	&    	\textemdash	&    	mfe	&    	10	\\
{\tt Info-RNA}        	&    	\checkmark	&    	\checkmark	&    	\textemdash	&    	\textemdash	&    	\textemdash	&    	\textemdash	&    	\textemdash	&    	'04	&    	1	&    	IUPAC	&    	\textemdash	&    	\textemdash	&    	mfe, prob	&    	50	\\
{\tt NUPACK}  	&    	\checkmark	&    	\checkmark	&    	\textemdash	&    	\checkmark$\star$	&    	\textemdash	&    	\textemdash	&    	\checkmark	&    	'99,'04	&    	0,1,2	&    	\checkmark	&    	\textemdash	&    	\textemdash	&    	ens def	&    	10	\\
{\tt MODENA}  	&    	\checkmark	&    	\textemdash	&    	\checkmark	&    	\textemdash	&    	\textemdash	&    	\textemdash	&    	\textemdash	&    	I	&    	def	&    	\textemdash	&    	\textemdash	&    	\textemdash	&    	mfe, prob	&    	?	\\
{\tt Frnakenstein}	&    	\checkmark	&    	\textemdash	&    	\textemdash	&    	\textemdash	&    	\checkmark	&    	\textemdash	&    	\checkmark	&    	I	&    	def	&    	\textemdash	&    	\textemdash	&    	\textemdash	&    	various	&    	?	\\
{\tt IncaRNAtion}	&    	\checkmark	&    	\textemdash	&    	\textemdash	&    	\textemdash	&    	\textemdash	&    	\textemdash	&    	\checkmark	&    	'04$\star$	&    	\textemdash	&    	IUPAC	&    	\textemdash	&    	\textemdash	&    	pf sampling	&    	\textemdash	\\
{\tt ERD}	&    	\checkmark	&    	\checkmark	&    	\textemdash	&    	\textemdash	&    	\textemdash	&    	\textemdash	&    	\checkmark	&    	I	&    	def	&    	IUPAC$\star$	&    	\textemdash	&    	\textemdash	&    	mfe	&    	MAX$\star$	\\
{\tt RNAdesign}	&    	\checkmark	&    	\textemdash	&    	\textemdash	&    	\textemdash	&    	\checkmark	&    	\textemdash	&    	\checkmark	&    	'04	&    	def	&    	\textemdash	&    	\textemdash	&    	\textemdash	&    	various	&    	\textemdash	\\
{\tt RNAfbinv}	&    	\checkmark	&    	\textemdash	&    	\textemdash	&    	\textemdash	&    	\textemdash	&    	\checkmark	&    	\textemdash	&    	'99, I	&    	def	&    	local A,C,G,U	&    	\textemdash	&    	\textemdash	&    	mfe	&    	\textemdash	\\
\hline
\end{tabular}
\caption{Comparison table for RNA inverse folding software. 
Column headers: Soft (Software), $\Downarrow$ (software can be downloaded),
WS (web server), PK (pseudoknots),
H (hybridization), MT (multiple targets), PT (partial targets),
T (temperature),EM (energy model), D (dangles), 
SeqC (sequence constraints),
StrC (structural constraints), AaC (amino acid constraints),
O (objective), Num (maximum number of sequences returned). 
{\em Comments:} In column H,
{\tt RNAiFold 2.0} and {\tt NUPACK} are the only programs that solve
inverse folding for target {\em hybridizations}; moreover, 
{\tt NUPACK} has `$\checkmark\star$', since it is the only
algorithm that allows hybridization of more than 2 strands.
In column EM, values are '99 (Turner'99), '04 (Turner'04),
'04$\star$ (Turner'04
base stacking parameters with no entropic free energies),
I (installed, depending on the version of Vienna RNA Package installed
on user's computer).
In column D, dangle status is 0 (no dangle), 1 (max of $5'$ and $3'$-dangle),
2 (sum of $5'$ and $3'$-dangle), 3 (dangles and coaxial stacking), def
(depending on default setting of user's version of Vienna RNA Package).
In column SeqC, values are \checkmark (IUPAC plus additional constraints)
IUPAC, IUPAC$\star$ 
(limited subset of IUPAC symbols), and local A,C,G,U (oligonucleotide
specified at a given position using only A,C,G,U). 
In column O, values are mfe (minimum free energy structure), 
prob (maximize Boltzmann probability), ens def (ensemble defect), 
pf sampling (partition function sampling with a restriction of Turner'04).
In column Num, the number of solutions returned by the web server is given
(\textemdash if no web server available); a question mark in this column
appears for {\tt MODENA} and {\tt Frnakenstein}, which are genetic algorithms,
and have a population of evolving sequences, so the user cannot request
a fixed number of solutions.
{\tt ERD} contains MAX$\star$, since the web server allows the user to
request an arbitrary number of {\em iterations} (distinct runs) of the
program, where 10 minutes is the maximum computation time alotted per request.
In contrast, {\tt RNAiFold 2.0} contains MAX in this column, which
indicates that as many solutions are returned as
possible within the system-dependent run time bound.
}
\label{table:1}
\end{table*}

\begin{table*}
\small
\begin{tabular}{|l||r|r|r|r|r|r|r|r|r|r|}
\hline
{\footnotesize Method}	&	{\footnotesize ERD}	&	{\footnotesize FRNA}	&	{\footnotesize Incarnation}	&	{\footnotesize Info-RNA}	&	{\footnotesize MODENA}	&	{\footnotesize Nupack}	&	{\footnotesize RNA-SSD}	&	{\footnotesize RNAfbinv}	&	{\footnotesize RNAiFold2}	&	{\footnotesize RNAinverse}	\\
\hline
{\footnotesize Output (\%)}	&	{\footnotesize 100\%}	&	{\footnotesize 30\%}	&	{\footnotesize 60\%}	&	{\footnotesize 95\%}	&	{\footnotesize 60\%}	&	{\footnotesize 57\%}	&	{\footnotesize 90\%}	&	{\footnotesize 13\%}	&	{\footnotesize 65\%}	&	{\footnotesize 65\%} \\
{\footnotesize Target (\%)}	&	{\footnotesize 85\%}	&	{\footnotesize 38\%}	&	{\footnotesize 0\%}	&	{\footnotesize 57\%}	&	{\footnotesize 45\%}	&	{\footnotesize 70\%}	&	{\footnotesize 82\%}	&	{\footnotesize 0\%}	&	{\footnotesize 100\%}	&	{\footnotesize 18\%} \\
{\footnotesize Avg str len}	&	{\footnotesize 397}	&	{\footnotesize 122}	&	{\footnotesize 352}	&	{\footnotesize 393}	&	{\footnotesize 234}	&	{\footnotesize 256}	&	{\footnotesize 400}	&	{\footnotesize 74}	&	{\footnotesize 363}	&	{\footnotesize 208} \\
{\footnotesize Avg output}	&	{\footnotesize 117}	&	{\footnotesize 325}	&	{\footnotesize 41,535}	&	{\footnotesize 195}	&	{\footnotesize 50}	&	{\footnotesize 22}	&	{\footnotesize 1}	&	{\footnotesize 2}	&	{\footnotesize 55,476}	&	{\footnotesize 935} \\
{\footnotesize P(S)}	&	{\footnotesize 3.32\%}	&	{\footnotesize 1.70\%}	&	{\footnotesize 0.06\%}	&	{\footnotesize 3.17\%}	&	{\footnotesize 11.30\%}	&	{\footnotesize 30.01\%}	&	{\footnotesize 2.24\%}	&	{\footnotesize 0.36\%}	&	{\footnotesize 23.21\%}	&	{\footnotesize 0.78\%} \\
{\footnotesize Native cont. (\%)}	&	{\footnotesize $85\pm 9$}	&	{\footnotesize $61\pm 15$}	&	{\footnotesize $63\pm 13$}	&	{\footnotesize $76\pm 12$}	&	{\footnotesize $89\pm 9$}	&	{\footnotesize $98\pm 1$}	&	{\footnotesize $85$}	&	{\footnotesize $32\pm 6$}	&	{\footnotesize $93\pm 2$}	&	{\footnotesize $57\pm 12$}\\
\hline \hline 
{\footnotesize Avg E}	&	{\footnotesize -0.41}	&	{\footnotesize -0.24}	&	{\footnotesize -0.46}	&	{\footnotesize -0.63}	&	{\footnotesize -0.46}	&	{\footnotesize -0.44}	&	{\footnotesize -0.30}	&	{\footnotesize -0.14}	&	{\footnotesize -0.56}	&	{\footnotesize -0.23} \\
{\footnotesize Pos entropy}	&	{\footnotesize 0.33}	&	{\footnotesize 0.71}	&	{\footnotesize 0.41}	&	{\footnotesize 0.44}	&	{\footnotesize 0.15}	&	{\footnotesize 0.07}	&	{\footnotesize 0.36}	&	{\footnotesize 0.88}	&	{\footnotesize 0.12}	&	{\footnotesize 0.80} \\
{\footnotesize MH diversity}	&	{\footnotesize 0.16}	&	{\footnotesize 0.35}	&	{\footnotesize 0.21}	&	{\footnotesize 0.22}	&	{\footnotesize 0.07}	&	{\footnotesize 0.03}	&	{\footnotesize 0.18}	&	{\footnotesize 0.45}	&	{\footnotesize 0.06}	&	{\footnotesize 0.38} \\
{\footnotesize Vienna diversity}	&	{\footnotesize 0.11}	&	{\footnotesize 0.23}	&	{\footnotesize 0.15}	&	{\footnotesize 0.16}	&	{\footnotesize 0.05}	&	{\footnotesize 0.02}	&	{\footnotesize 0.11}	&	{\footnotesize 0.30}	&	{\footnotesize 0.05}	&	{\footnotesize 0.26} \\
{\footnotesize Exp bp dist}	&	{\footnotesize 0.09}	&	{\footnotesize 0.21}	&	{\footnotesize 0.27}	&	{\footnotesize 0.16}	&	{\footnotesize 0.06}	&	{\footnotesize 0.01}	&	{\footnotesize 0.08}	&	{\footnotesize 0.38}	&	{\footnotesize 0.03}	&	{\footnotesize 0.24} \\
{\footnotesize Ens def}	&	{\footnotesize 0.14}	&	{\footnotesize 0.32}	&	{\footnotesize 0.39}	&	{\footnotesize 0.22}	&	{\footnotesize 0.08}	&	{\footnotesize 0.02}	&	{\footnotesize 0.14}	&	{\footnotesize 0.56}	&	{\footnotesize 0.04}	&	{\footnotesize 0.37} \\
{\footnotesize Exp num bp}	&	{\footnotesize 0.28}	&	{\footnotesize 0.29}	&	{\footnotesize 0.34}	&	{\footnotesize 0.30}	&	{\footnotesize 0.26}	&	{\footnotesize 0.29}	&	{\footnotesize 0.28}	&	{\footnotesize 0.28}	&	{\footnotesize 0.27}	&	{\footnotesize 0.28} \\
{\footnotesize GC-content (\%)}	&	{\footnotesize $55\%$}	&	{\footnotesize $49\%$}	&	{\footnotesize $71\%$}	&	{\footnotesize $72\%$}	&	{\footnotesize $50\%$}	&	{\footnotesize $57\%$}	&	{\footnotesize $36\%$}	&	{\footnotesize $51\%$}	&	{\footnotesize $57\%$}	&	{\footnotesize $49\%$}\\
\hline
\end{tabular}
\caption{\small Comparison of 10 programs for RNA inverse folding, benchmarked on 63 target structures, as explained in the text. Averages are
given, rounded either to two decimals or to the nearest integer as appropriate.
Complete data, with averages and standard deviations, can be found on the
web server {\tt RNAiFold 2.0}. FRNA stands for {\tt FRNAnkenstein}.
Row labels are as follows, whereby measures appearing after the double line have been normalized by dividing by sequence length -- for instance,
{\em Avg E} denotes the {\em normalized} average free energy of the returned sequences, computed as the average, taken over all 63 individual target
structures $S_0$, of average normalized free energies
${E(\aseq,S_0)}/{|\aseq|}$, taken over all sequences 
$\aseq$ returned for target structure $S_0$, where $E(\aseq,S_0)$ denotes the free energy of sequence $\aseq$
with respect to the structure $S_0$. The other normalized measures are defined in an analogous manner.
{\em (Unnormalized measures)}
Output (\%): Fraction of the 63 target structures for which some output was produced.
Target (\%): Average fraction of output sequences whose MFE structure is the target.
Avg str len: Average target structure length, taken over those target structures for which at least one output sequence was returned.
Avg output: Total number of sequences returned for all 63 targets, divided by the number of targets for which at least one sequence was returned.
P(S): average probability of target structure, defined as the average,
taken over all 63 target structures $S_0$,
of the average Boltzmann probability $P(s,S_0) = ({\exp(-E(s,S_0)/RT})/{Z}$,
taken over all sequences $s$ returned for target structure $S_0$.
{\em (Normalized measures)}
Avg E: normalized average free energy with respect to target (previously
defined). The remain measures are length-normalized versions of 
positional entropy,
Morgan-Higgs diversity,
Vienna diversity,
expected base pair distance from target structure,
ensemble defect with respect to target structure,
expected number of base pairs,
proportion of native contacts, and
GC-content. Measures are defined in the text.
}
\label{table:comparison}
\end{table*}

\hfill\break\newpage

\begin{figure*}
\begin{subfigure}[b]{0.46\textwidth}
\includegraphics[width=\textwidth]{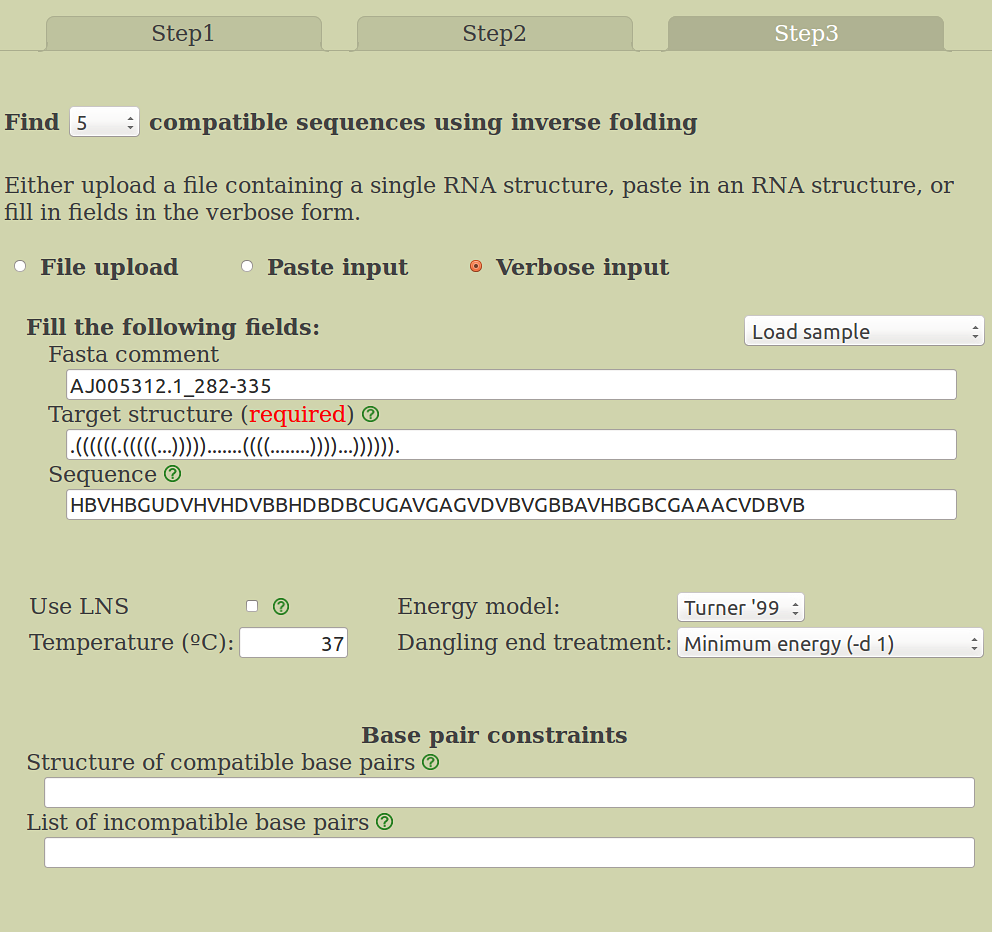}
\caption{}
\label{fig:A2}
\end{subfigure}%
\hskip  0.04\textwidth
\begin{subfigure}[b]{0.46\textwidth}
\includegraphics[width=\textwidth]{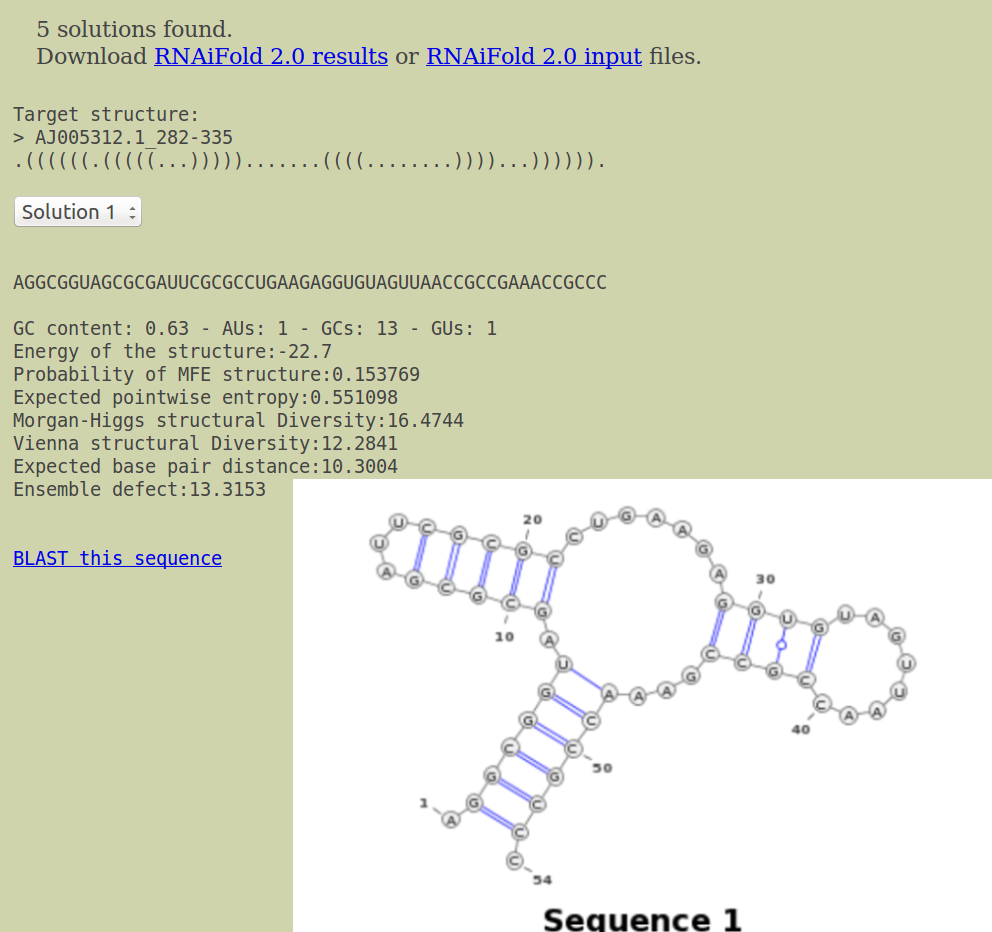}
\caption{}
\label{fig:B2}
\end{subfigure}
\caption{
{\em (a)}
Third of 3 screens in {\em Rfam-based Design}, where the user
can enter additional structure compatibility and incompatibility constraints,
which require all solutions to be compatible with a second structure
(in addition to folding into the target structure), and which do not
allow base pairing at positions stipulated in the incompatibility constraints.
{\em (b)} Output from the pipeline described in the three previous screen
shots. Note that the GC-content, average positional entropy,
ensemble defect and other structural diversity measures are computed. These
measures provide an idea of how similar the low energy ensemble of structures
resembles the minimum free energy structure, which is guaranteed to be 
identical to the user-input target structure.
}
\label{fig:screenShotDesign3-4}
\end{figure*}

\begin{figure*}
\begin{subfigure}[b]{0.46\textwidth}
\includegraphics[width=\textwidth]{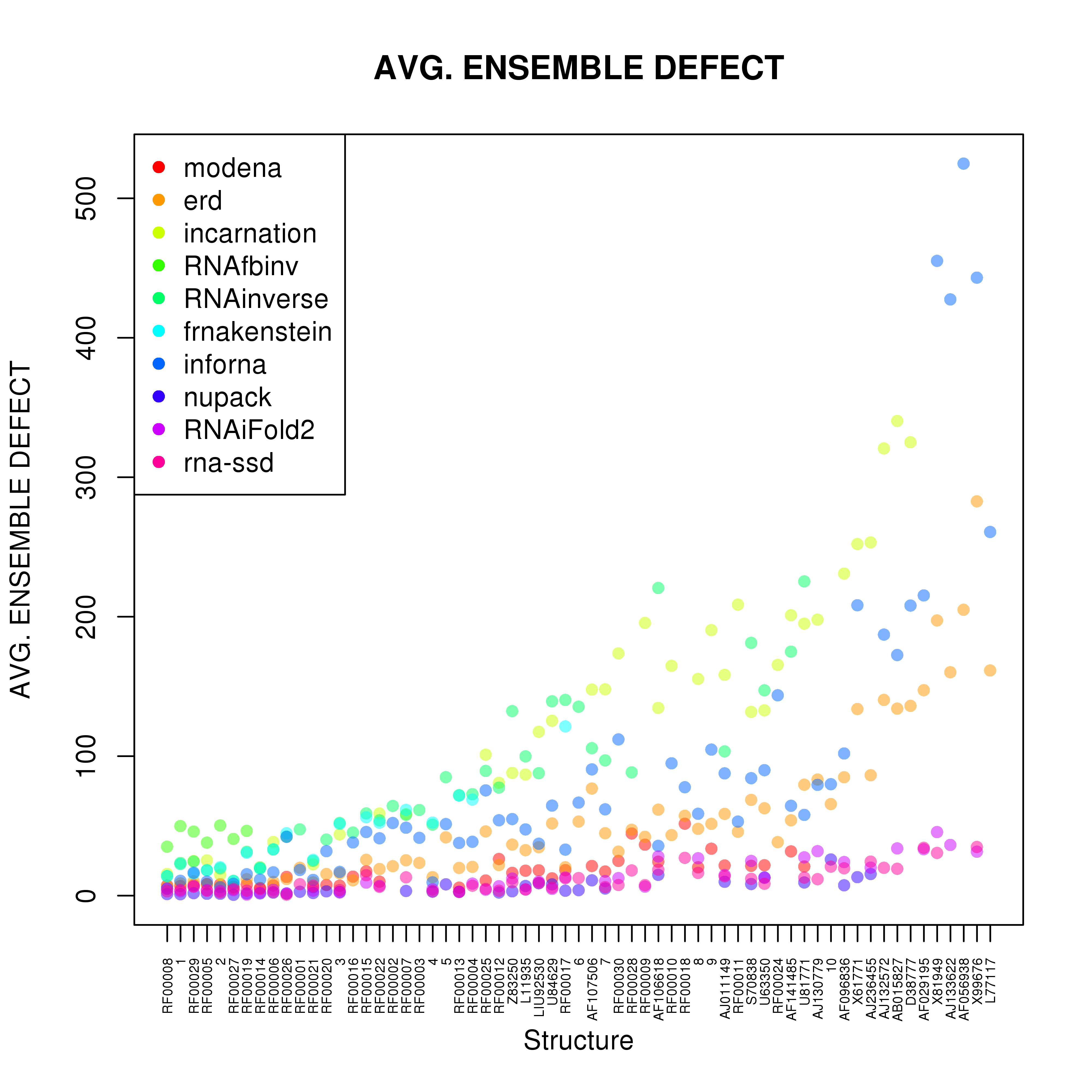}
\caption{}
\label{fig:ensDef}
\end{subfigure}%
\hskip  0.04\textwidth
\begin{subfigure}[b]{0.46\textwidth}
\includegraphics[width=\textwidth]{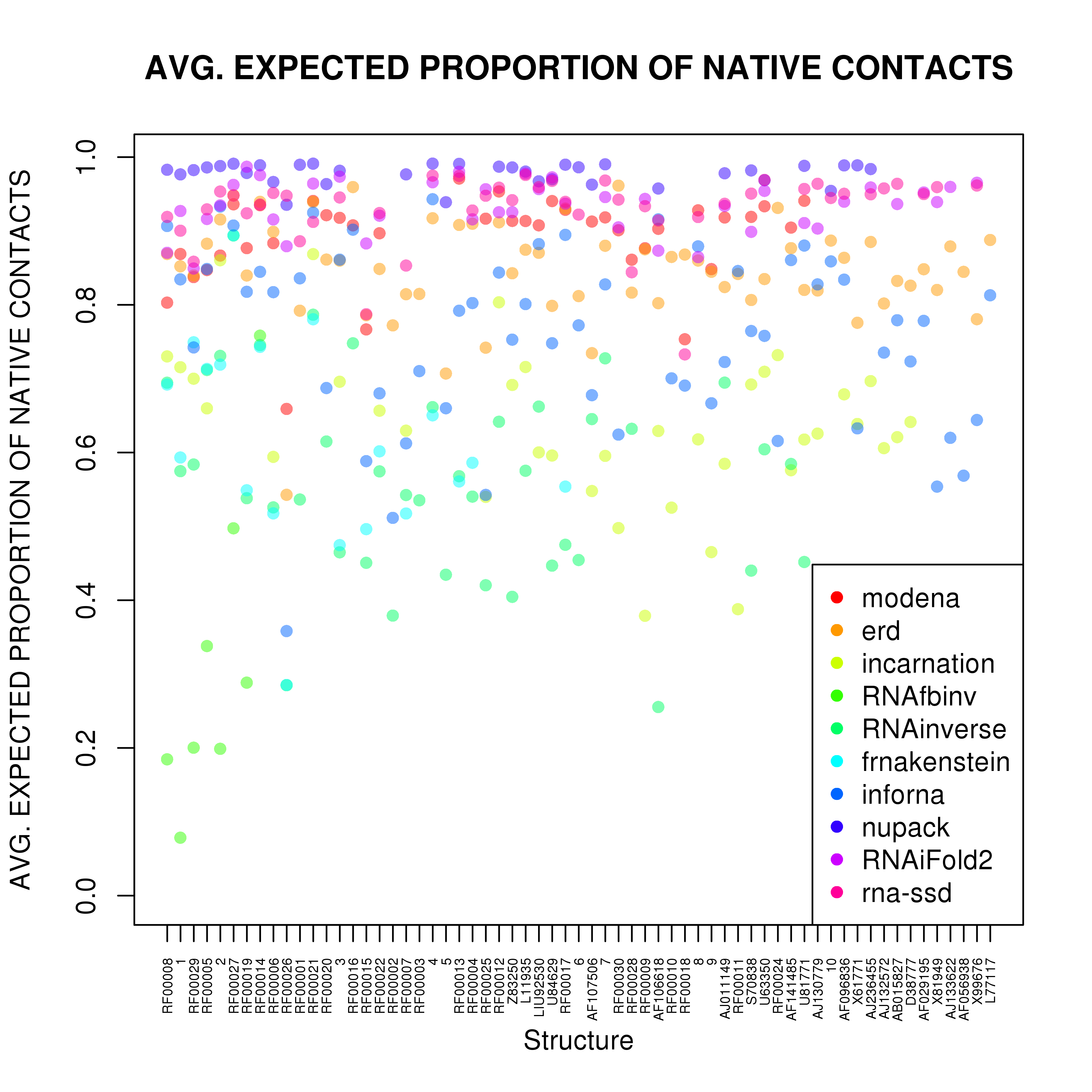}
\caption{}
\label{fig:posEntropy}
\end{subfigure}
\caption{
{\em (a)}
Average ensemble defect for inverse folding software, where the
(non length-normalized) averages are taken over all sequences returned
for a fixed target structure. Name of the target structure is given
on the $x$-axis, arranged in increasing length (length is not drawn
to scale); $y$-axis depicts the average ensemble defect for the output
of each software, on each target.
{\em (b)} Average proportion of native (target) contacts,
where the
(non length-normalized) averages are taken over all sequences returned
for a fixed target structure. Name of the target structure is given
on the $x$-axis, arranged in increasing length (length is not drawn
to scale); $y$-axis depicts the expected proportion of base pairs in the 
target structure that are present in the low energy ensemble of all
structures, for each target.  Benchmarking data, both raw data and 
length-normalized data, as well as scatter plots for
for all measures can be found at the {\tt RNAiFold 2.0} web site in
the `Download' tab.
}
\label{fig:structuralDiversity}
\end{figure*}

\end{document}